\documentclass[fleqn,usenatbib]{mnras}

\usepackage{newtxtext,newtxmath}

\usepackage[T1]{fontenc}

\DeclareRobustCommand{\VAN}[3]{#2}
\let\VANthebibliography\thebibliography
\def\thebibliography{\DeclareRobustCommand{\VAN}[3]{##3}\VANthebibliography}

\usepackage{graphicx}	
\usepackage{amsmath}	

\newcommand{\thirteenco}{\mbox{$^{13}$CO}} 
\newcommand{\twelvecoh}{\mbox{$^{12}$CO($J$ = 2--1)}} 
 
\newcommand{\thirteencoh}{\mbox{$^{13}$CO($J$ = 2--1)}} 
\newcommand{\twelvecol}{\mbox{$^{12}$CO($J$ = 1--0)}}

\newcommand{\msun}{\mbox{$M_\odot$}}
\newcommand{\kms}{\mbox{km~s$^{-1}$}}
\newcommand{\kkms}{\mbox{K~km~s$^{-1}$}}

\newcommand{\hone}{\mbox{H{\sc i}}}
\newcommand{\htwo}{\mbox{H{\sc ii}}}

\title[Cloud-cloud collision in Sh2-233]{Evidence for a Cloud-Cloud Collision in Sh2-233 Triggering the Formation of the High-mass Protostar Object IRAS~05358+3543}

\author[R. Yamada et al.]{
Rin I. Yamada,$^{1}$\thanks{E-mail: yamada@a.phys.nagoya-u.ac.jp}
Yasuo Fukui,$^{2}$
Hidetoshi Sano,$^{3}$
Kengo Tachihara,$^{1}$
John H. Bieging,$^{4}$
Rei Enokiya,$^{5}$
\newauthor
Atsushi Nishimura,$^{6}$
Shinij Fujita,$^{1,7}$
Mikito Kohno,$^{8}$
and
Kisetsu Tsuge,$^{1}$
\\
$^{1}$Department of Physics, Nagoya University, Furo-cho, Chikusa-ku, Nagoya 464-8601, Japan\\
$^{2}$Institute for Advanced Research, Nagoya University, Furo-cho, Chikusa-ku, Nagoya 464-8601, Japan\\
$^{3}$National Astronomical Observatory of Japan, Mitaka, Tokyo 181-8588, Japan\\
$^{4}$Steward Observatory, The University of Arizona, Tucson, AZ 85721, USA\\
$^{5}$Department of Physics, Faculty of Science and Technology, Keio University, 3-14-1 Hiyoshi, Kohoku-ku,Yokohama, Kanagawa 223-8522, Japan\\
$^{6}$Institute of Astronomy, the University of Tokyo, 2-21-1 Osawa, Mitaka, Tokyo 181-0015, Japan\\
$^{7}$Department of Physical Science, Graduate School of Science, Osaka Prefecture University, 1-1 Gakuen-cho, Naka-ku, Sakai 599-8531, Japan\\
$^{8}$Astronomy Section, Nagoya City Science Museum, 2-17-1 Sakae, Naka-ku, Nagoya, Aichi, 460-0008, Japan
}

\date{Accepted XXX. Received YYY; in original form ZZZ}

\pubyear{2015}
\begin{document}
\label{firstpage}
\pagerange{\pageref{firstpage}--\pageref{lastpage}}
\maketitle

\begin{abstract}
We have carried out a new kinematical analysis of the molecular gas in the Sh2-233 region by using the CO $J$~=~2--1 data taken at $\sim$0.5~pc resolution. The molecular gas consists of a filamentary cloud of 5-pc length with 1.5-pc width where two dense cloud cores are embedded. The filament lies between two clouds, which have a velocity difference of 2.6~$\kms$ and are extended over $\sim$5 pc. We frame a scenario that the two clouds are colliding with each other and compressed the gas between them to form the filament in $\sim$0.5 Myr which is perpendicular to the collision. It is likely that the collision formed not only the filamentary cloud but also the two dense cores. One of the dense cores is associated with the high-mass protostellar candidate IRAS~05358+3543, a representative high-mass protostar. In the monolithic collapse scheme of high mass star formation, a compact dense core of 100~$\msun$ within a volume of 0.1~pc radius is assumed as the initial condition, whereas the formation of such a core remained unexplained in the previous works. We argue that the proposed collision is a step which efficiently collects the gas of 100~$\msun$ into 0.1~pc radius. This lends support for that the cloud-cloud collision is an essential process in forming the compact high-mass dense core, IRAS~05358+3543.
\end{abstract}



\section{Introduction}\label{sec:intro}
High-mass star formation is an important process which substantially affects the galaxy evolution via 
enormous energy inputs into the interstellar medium (ISM). The energy inputs regulate the physical 
conditions of the ISM and control the gravitational collapse of the ISM which leads to high-mass star 
formation. The mechanism of high-mass star formation, however, has been an issue difficult to solve. 
The reason for this is multi-fold. The region of high-mass star formation is rare in the solar 
neighbourhood and detailed observations of this process were difficult as compared with the sites of low-mass 
star formation. Further, the timescale of high-mass star formation is probably very short, less than Myr, 
because of higher density of the star forming core, making a marked contrast with the low-mass star formation which 
proceeds slowly in 100~Myr due to low density. Because of this contrast, there is a high barrier in finding out the high-mass star formation location and to resolve the star formation process therein with sufficient details.
In spite of such difficulties, progress has been achieved by extensive surveys for massive dense cores, and hundreds of such cores have been catalogued in the 
Galactic plane (e.g., \citealp{Beuther_catalogue, ATLASGAL_catalogue}). Massive dense cores are believed to be precursors of high-mass stars 
because a high mass protostar should experience a massive dense core phase prior to a mature high-mass 
star. It is however not certain if all the massive cores are actual sites of high-mass star formation, 
since most of them lack a $\htwo$ region, an unmistakable sign of high-mass stars emitting strong UV
radiation. It was also puzzling that the number of such cores are large, suggesting a long timescale 
like 10 Myr, leaving room to suspect that many of the cores, even if not all, may not be such precursors 
but stay stable as they are.

The observational difficulties may be overcome by theoretical works. \cite{Mckee_and_tan} investigated typical 
density inherent to the earliest phase of the high mass star formation in the literature, and found that many
of the high mass star forming regions have common surface mass density around 1 g cm$^{-2}$, which 
corresponds to column density of $\sim 3\times10^{23}$~cm$^{-2}$. Cases of the high-column density in high-mass star formation 
include the massive clusters, the Orion Nebula Cluster in M42 and W3~Main, where around ten O stars are 
formed (for reviews \citealp{Odel} and \citealp{2008hsf1.book..264M}), as well as a more isolated high-mass 
protostar IRAS~05358+3543 in Sh2-233 (\citealp{Beuther2007}; see for a review \citealp{Reipurth2008}). 
\cite{Krumholz_mass_acc} made hydrodynamical numerical simulations of gravitational collapse of a massive 
cloud core under the assumed initial condition, 100 $\msun$ within 0.1 pc, corresponding to the condition 
found in the above regions. The simulations, which are usually referred to as the monolithic collapse 
models, successfully produced two $\sim$30~$\msun$ stars in binary in a timescale of a few times 10$^4$ yr, and 
showed that the issue of high mass star formation may be solved by the model. 

It however remained to be explained how such a massive dense core can be formed in the ISM. The fact that such massive dense cores can form high-mass stars rapidly without fail suggests that the core may rapidly collapse to lower-mass stars well before collected as a massive core. If one traces the whole gravitational collapse initiating by low density like 1000~cm$^{-3}$, the initial condition assumed 
by the monolithic collapse model might be hardly realised with self-gravity alone. This reasoning may be 
supported by the scenarios of a cloud-cloud collision in a number of star forming regions by the recent observational works including well known $\htwo$ regions M20, M42 and M17 \citep{Torii2017, Fukui_2018_Orion, Nishimura_M17}, giant molecular clouds with mini-starbursts W43, W51, and Carina \citep{Kohno_W43, Fujita_w51, Fujita_carina}, the Galactic centre \citep{Enokiya_2021_compile, Tsuboi2021}, the Magellanic Clouds \citep{Fukui_magellan, Tsuge2019}, M33 \citep{Sano_2021_M33, Tokuda2020, Kondo_2021}, and the Antennae galaxies \citep{Tsuge_antennae, Tsuge_B1} as well as
theoretical results of colliding molecular gas flows \citep{Inoue_Fukui, Inoue_2018, Fukui_core_mass}. See also \cite{Fukui_rev} for a review of the related observational and theoretical works.
Such colliding gas flows can significantly accelerate the gas collection by the supersonic motion, ten 
times faster than the free fall, without converting gas into stars, and may be a crucial step in forming a 
massive core.

In the present paper, we intend to make a detailed kinematic study of the gas in the region of Sh2-233IR at $\sim$10\arcmin ~to the southeast of Sh2-233, one of the representative cases of high gas surface density including the high mass protostar candidate IRAS~05358+3543. The region harbours other $\htwo$ regions Sh2-231, Sh2-232, and Sh2-235 in addition to Sh2-233 (see for a review \citealp{Reipurth2008}), which are located at a distance of 1.8~kpc \citep{evans_blair} and within 
40~pc of Sh2-233 on its eastern side in the galactic coordinate. This is a relatively dense 
region of star formation along the Galactic plane. It may be considered that these $\htwo$ regions are 
expanding to accelerate the gas and triggering star formation in the surroundings 
according to a picture of \cite{Elmegreen_Lada}, because the $\htwo$ regions can agitate 
the surrounding gas to accelerate expanding motions in a timescale of several Myrs or 
more, the probable duration of these extended $\htwo$ regions. The active O star formation attracted 
attention and stimulated a number of works on star formation at multi-wavelength. 
Most recently, a detailed analysis of the CO gas in Sh2-235 were carried out by 
\cite{Dewangan2017} (see the references therein for observational works at multi-wavelength). This work suggests that a cloud cloud collision is a viable mechanism of a recent 
trigger of star formation within a timescale of Myr, whereas the acceleration of the gas 
clouds, which may be $\hone$ or H$_2$ clouds, could have originated more than a few Myr ago by acceleration due to $\htwo$ regions. 

The present work is organised as follows; Section \ref{sec:data} 
summarises the data used in the analysis and Section \ref{sec:results} the results of the analysis. Section \ref{sec:discussion} discusses the implications and Section \ref{sec:conclusion} gives conclusions.

\section{Datasets}\label{sec:data}
We made use of the archival $J$ = 2--1 data of the $^{12}$CO and $^{13}$CO emission 
obtained with the Heinrich Hertz Submillimeter Telescope (SMT) at the Arizona 
Radio Observatory \citep{Bieging2016}. Entire observations were carried 
out by using the fully-sampled on-the-fly (OTF) method with a full half power beam width (HPBW) of 
32$\arcsec$. The $^{12}$CO and $^{13}$CO data are both convolved to a spatial resolution of 38$\arcsec$. 
Both $^{12}$CO and $^{13}$CO data cubes were resampled onto identical velocity grids with 0.15 $\kms$ spacing by third-order interpolation. The typical noise fluctuations are 0.10~K for $^{12}$CO and 0.12~K for $^{13}$CO at 0.15~$\kms$ velocity grid. For more details of the observations, see \cite{Bieging2016}. 
\section{Results}\label{sec:results}
\subsection{Gas distribution toward the Sh2-233 region}
\begin{figure}
	\includegraphics[width=\columnwidth]{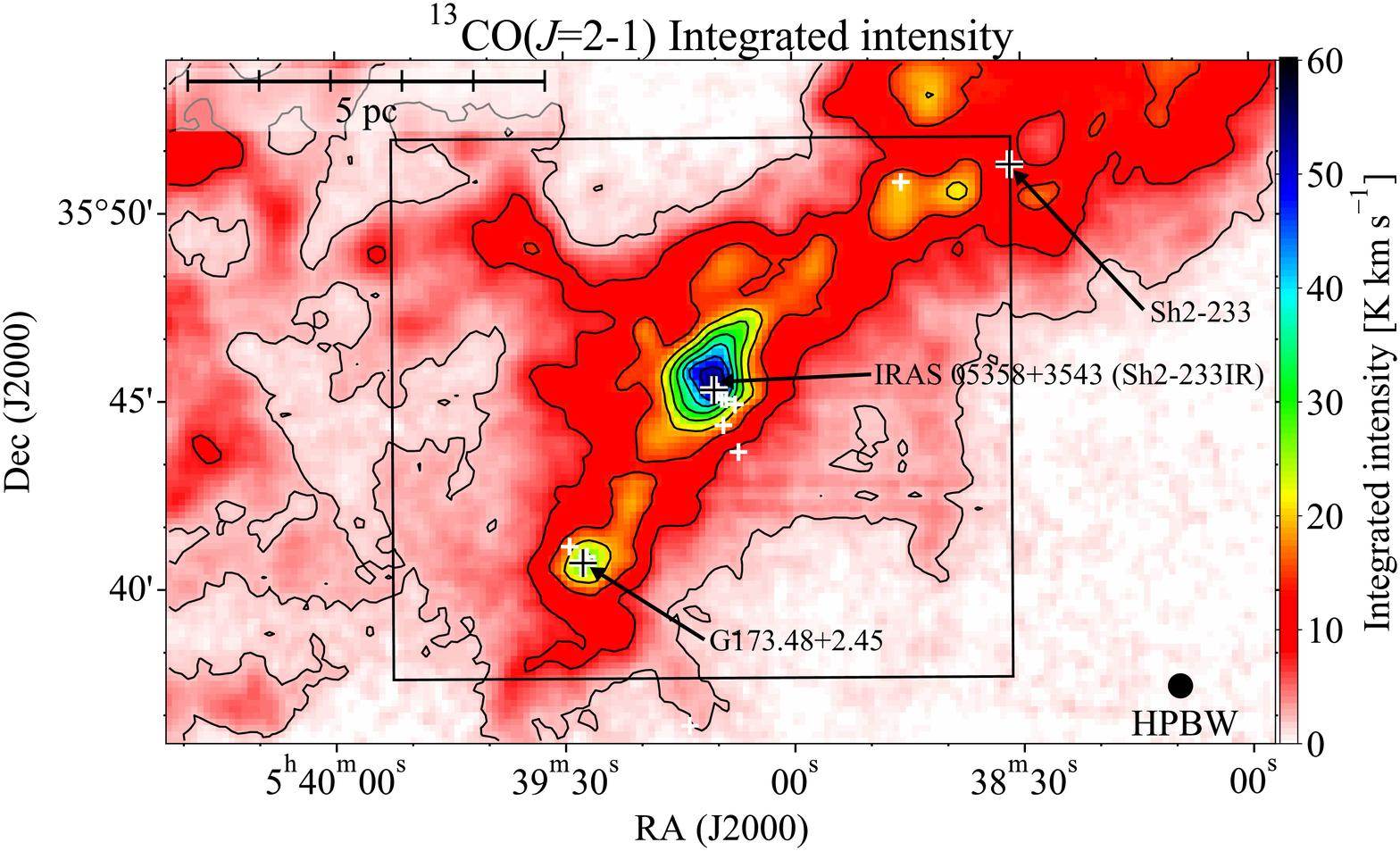}
  \caption{Integrated intensity distribution of the $\thirteencoh$ emission for a velocity range from 
  $-22$~$\kms$ to $-12$~$\kms$. Contours are plotted every 6~$\kkms$ from 2~$\kkms$. 
  Black crosses indicate the positions of Sh2-233, IRAS~05358+3543, and G173.58+2.45 corresponding to ($\alpha_\mathrm{J2000}$, $\delta_\mathrm{J2000}$) = ($05^\mathrm{h}38^\mathrm{m}31.5^\mathrm{s}$, $35$\degr51\arcmin19\arcsec), ($\alpha_\mathrm{J2000}$, $\delta_\mathrm{J2000}$) = ($05^\mathrm{h}39^\mathrm{m}10.4^\mathrm{s}$, $35$\degr45\arcmin19\arcsec), ($\alpha_\mathrm{J2000}$, and $\delta_\mathrm{J2000}$) = ($05^\mathrm{h}39^\mathrm{m}27.7^\mathrm{s}$, $35$\degr40\arcmin43\arcsec), respectively. White crosses show the positions of YSOs detected by \citet{Marton2016}. Black rectangle outlines the region for which we obtained masses in section \ref{sec:results}}
    \label{Integrated_intensity}
\end{figure}

The molecular gas in the Sh2-233 region has velocity in a range from $-22$~$\kms$ to $-12$~$\kms$. 
Figure \ref{Integrated_intensity} shows the total integrated intensity of $\thirteencoh$ in the velocity range. Hereafter, we shall use mostly $\thirteencoh$ which is optically thin and suited for tracing gas density. 
The molecular gas in the region shows filamentary distribution (hereafter ``filament'') 
extending from the southeast to the northwest. Toward the brightest $\thirteencoh$ peak at ($\alpha_\mathrm{J2000}$, $\delta_\mathrm{J2000}$) $\sim$ ($05^\mathrm{h}39^\mathrm{m}12^\mathrm{s}$, $35$\degr45\arcmin40\arcsec), the high-mass protostar candidate IRAS~05358+3543 shown by a black cross and a few additional YSOs are located as shown by the white crosses \citep{Marton2016}.
Another peak corresponding to G173.58+2.45 is located in the southeast at ($\alpha_\mathrm{J2000}$, $\delta_\mathrm{J2000}$) $\sim$ ($05^\mathrm{h}39^\mathrm{m}30^\mathrm{s}$, 
$35$\degr40\arcmin) and 
the third one at the northwest ($\alpha_\mathrm{J2000}$, $\delta_\mathrm{J2000}$) $\sim$ ($05^\mathrm{h}38^\mathrm{m}45^\mathrm{s}$, $35$\degr40\arcmin). 

\begin{figure*}
	\includegraphics[width=\textwidth]{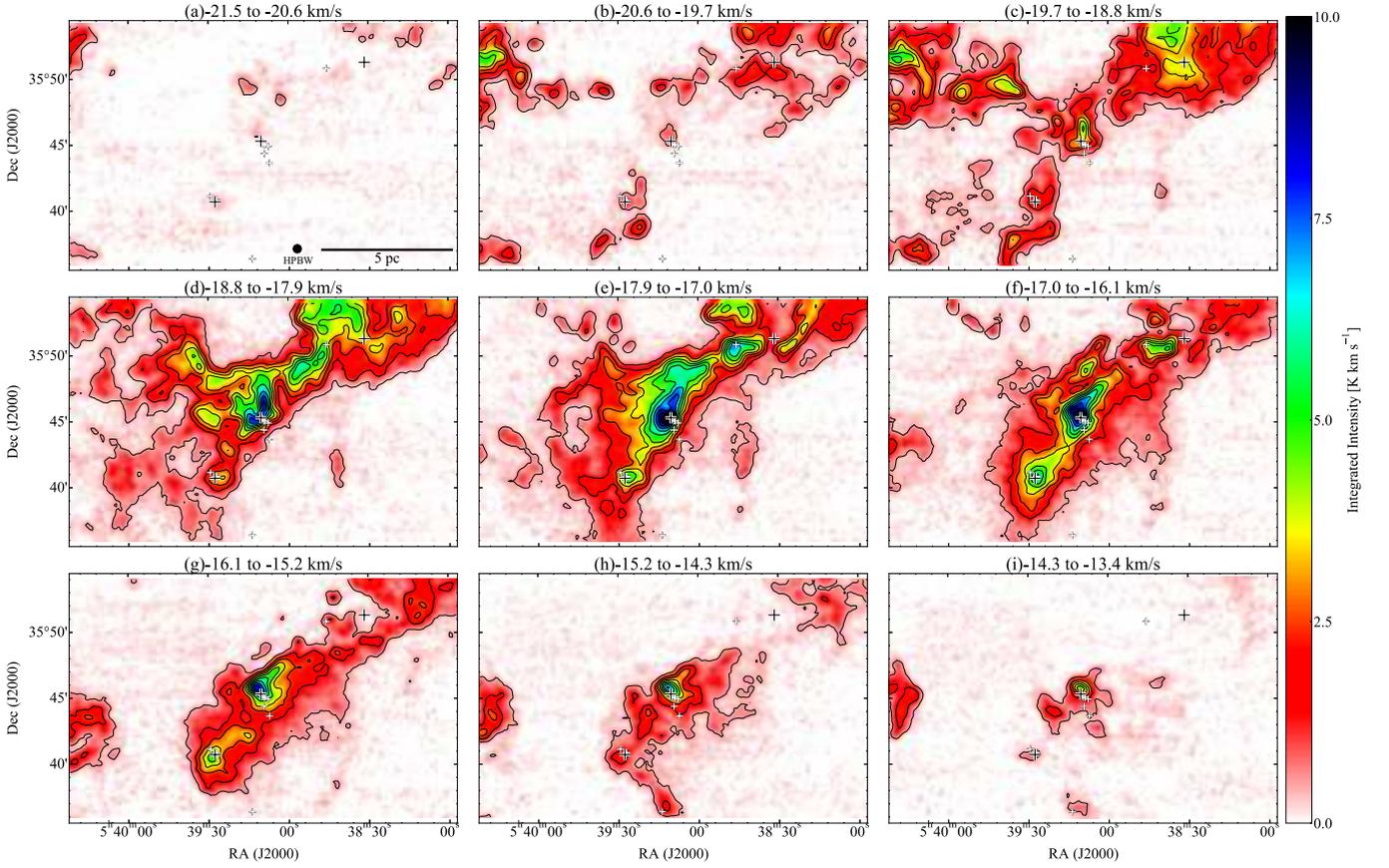}
  \caption{Velocity channel distribution of the $\thirteencoh$ line emission. Integration velocity ranges are denoted at the top of each panel. Contours are plotted every 1~$\kkms$ from 0.5~$\kkms$. Black crosses and white crosses are the same as Fig. \ref{Integrated_intensity}}
    \label{chmap}
\end{figure*}

Figure \ref{chmap} shows velocity channel distributions of $\thirteencoh$ ranging from $-21.5$~$\kms$ to $-13.4$~$\kms$ every 0.9~$\kms$. Many panels show the filament with local peaks as substructures. The Panels (a) to (e) show the blue-shifted component and the panel (g) to (i) show red-shifted component.

\begin{figure}
	\includegraphics[width=\columnwidth]{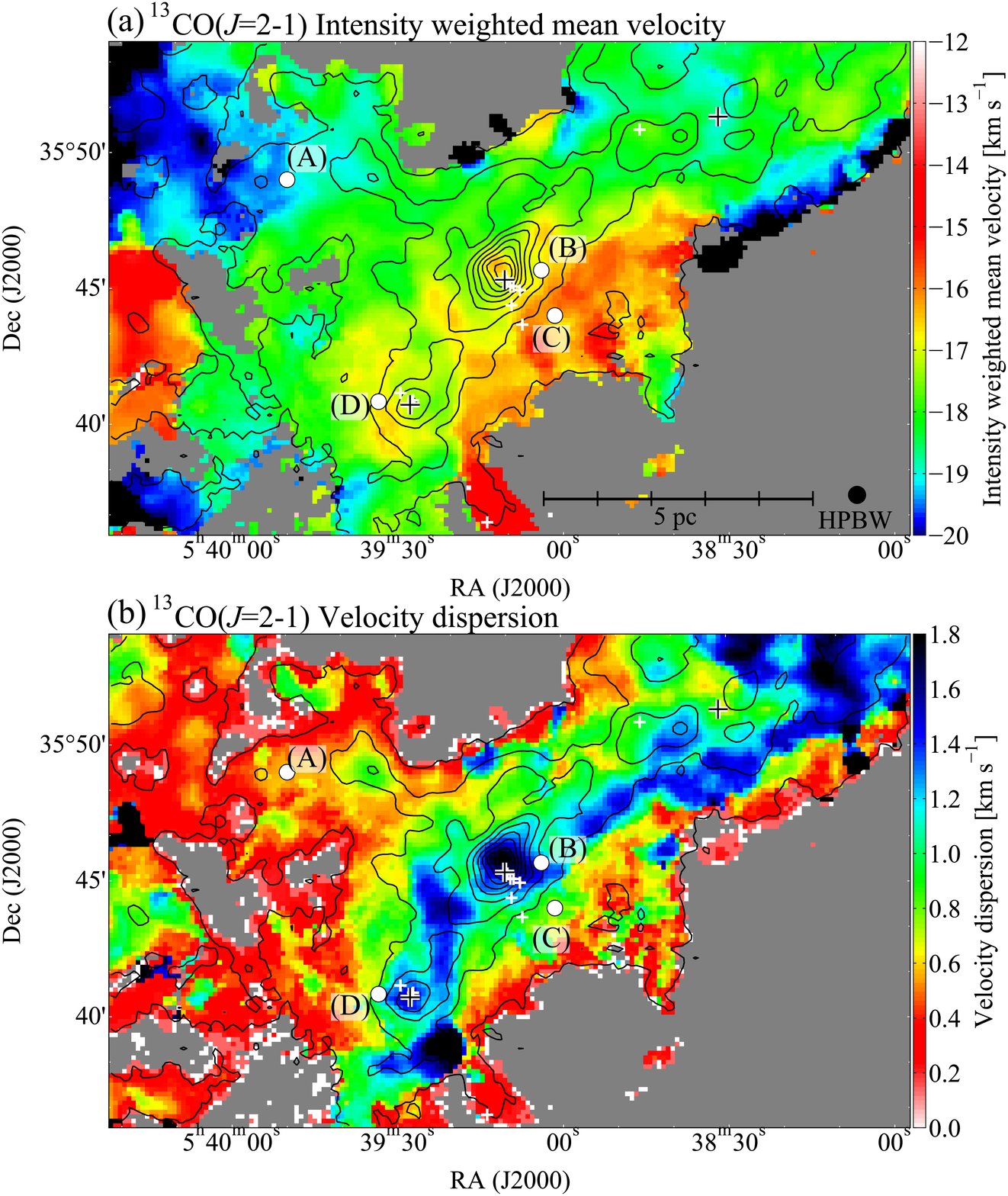}
  \caption{(a) Velocity centroid
  (the first moment) distribution of the 
  $\thirteencoh$ emission. The calculation velocity 
  range is from $-22$~$\kms$ to $-12$~$\kms$. Superposed
  contours show the integrated intensity distribution
  , same as the Fig. \ref{Integrated_intensity}. Black crosses and white 
  crosses indicate the positions of Sh2-233, 
  IRAS05358+3543, G173.58+2.45, and YSOs 
  detected by \citet{Marton2016}, respectively.
  White dots (A) to (C) indicates the position we present
  spectrum in figure \ref{spec}.
  (b) Velocity dispersion (the second moment) 
  distribution of the $\thirteencoh$ emission. Superposed
  contours are the same as (a). Black crosses and white crosses are the same as Fig. \ref{Integrated_intensity}}
    \label{moments}
\end{figure}

Figure \ref{moments}a shows the first moment of 
$\thirteencoh$ where integrated intensity >10 $\kkms$ is uniform 
at about $-16$~$\kms$ in the southwest, and that is 
uniform at around $-18$~$\kms$ in the northeast.
We find that 
there is a velocity difference of $\sim$2.6~$\kms$ between 
the southwestern and northeastern components 
within $\sim$5\arcmin~of the filament.
Figure \ref{moments}b shows the second moment of 
$\thirteencoh$, which is enhanced by a factor of 2--4 from 0.6~$\kms$ to 
1.6~$\kms$ toward the filament with a width of 1--2~pc.
It seems that the red-shifted component is somewhat 
more localised in the $\alpha_\mathrm{J2000}$ from $05^\mathrm{h}38^\mathrm{m}30^\mathrm{s}$ to $05^\mathrm{h}39^\mathrm{m}20^\mathrm{s}$
in panels (h) and (i) of Figure \ref{chmap} and Figure \ref{moments}a. 

\begin{figure}
	\includegraphics[width=\columnwidth]{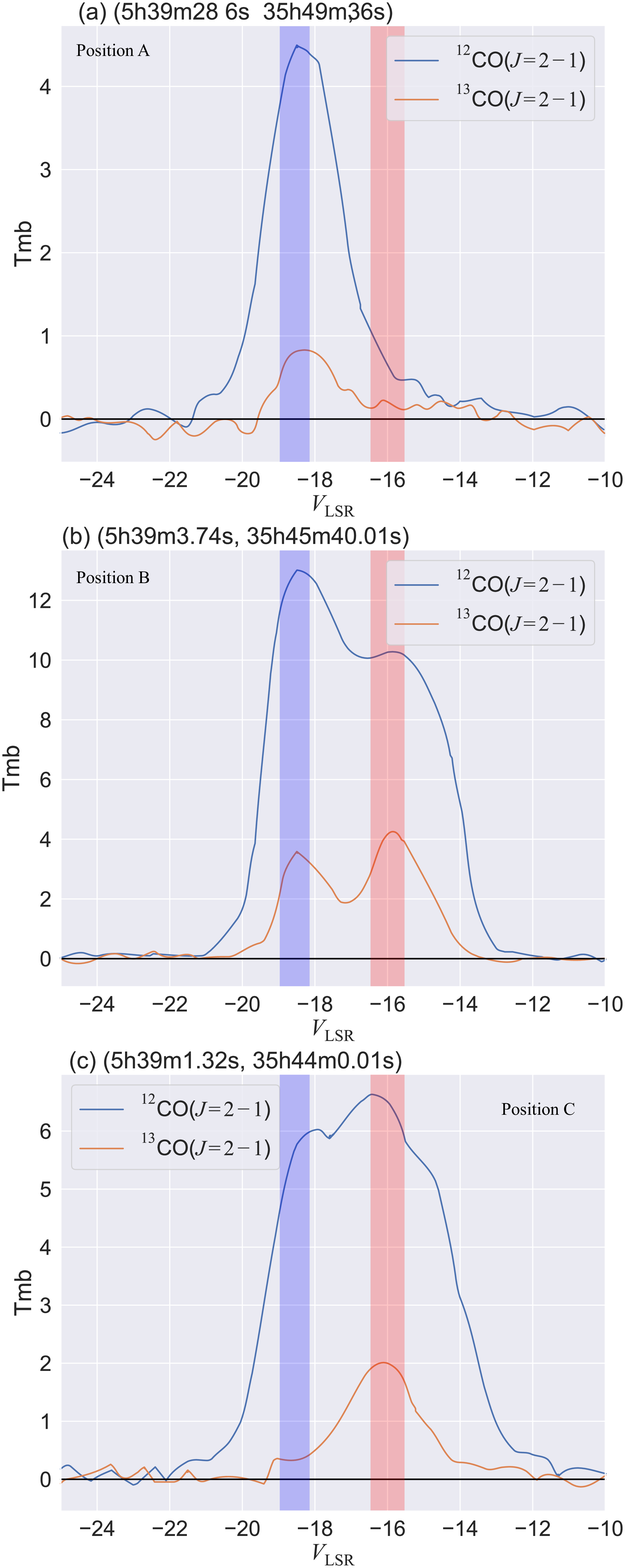}
  \caption{$\twelvecoh$ and $\thirteencoh$ spectra 
  at the position (a) to (c) in Fig. \ref{moments}. Red and blue bands show 
  the representative velocity ranges of each cloud defined in section \ref{sec:results}}
    \label{spec}
\end{figure}

Figure \ref{spec} shows typical CO profiles near Sh2-233IR in 
the three positions in Figure \ref{moments}a (A), (B), and (C). 
Figures \ref{spec}a and \ref{spec}c show single peaks which are blue-shifted in (a) and red-shifted in (c) and the peak 
velocity coincides with the first moment in Figure \ref{moments}a
, and Figure \ref{spec}b shows a double 
peak (B) in both $\twelvecoh$ and $\thirteencoh$.
Considering all of the above, we interpret that gas in the region of Sh2-233 consists of two different 
velocities toward the filament. We hereafter call the red-shifted component red cloud and the 
blue-shifted component blue cloud.

\begin{figure*}
	\includegraphics[width=\textwidth]{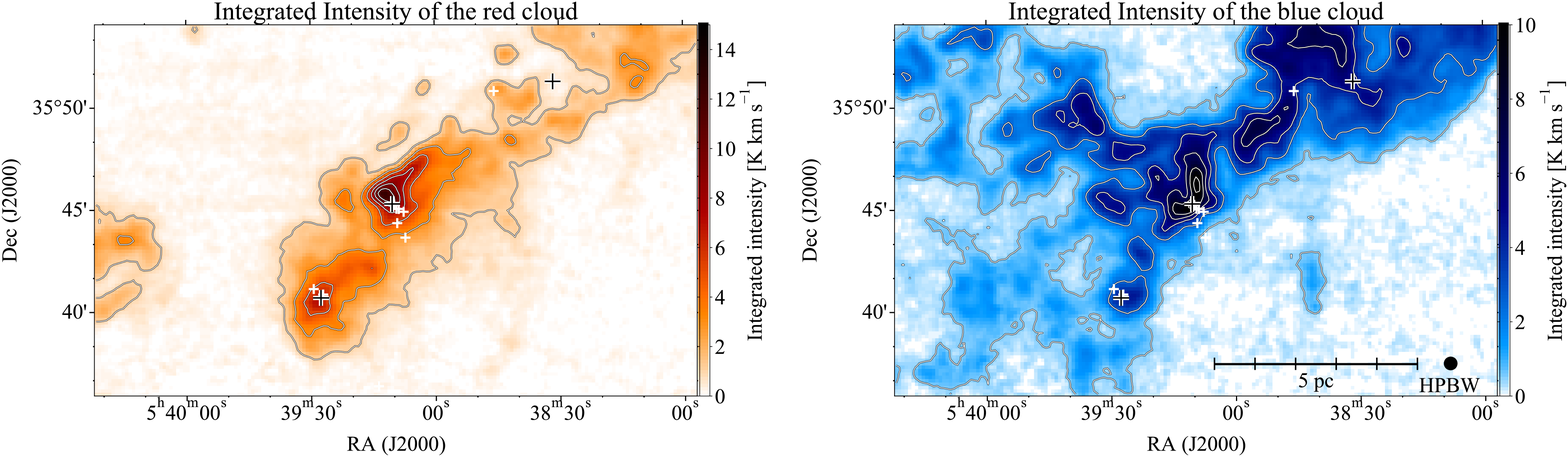}
  \caption{(a) Integrated intensity distribution of the red cloud. The integration range is from $-15.5$~$\kms$
  to $-16.5$~$\kms$. Contours are plotted every 2 $\kkms$ from 0.5~$\kkms$. (b) Integrated intensity 
  distribution of the blue cloud. The integration range is from $-19.0$~$\kkms$
  to $-18.15$~$\kkms$. The lowest contour and contour intervals are the same as (a).
  Black crosses and white crosses are the same as Fig. \ref{Integrated_intensity}}
    \label{twoclouds}
\end{figure*}

\subsection{The red- and blue-shifted clouds}
We now consider the properties of the red and blue-shifted clouds in Sh2-233. 
Following \cite{Enokiya_2021} 
who analysed two molecular clouds in the 
NGC~2024 region, we use a method based on the first moment 
and the second moment to define representative velocity ranges of the two clouds. 
First, we assume that the two clouds are overlapping 
toward the intermediate region with
non-overlapping outside the region. This means 
that the northeastern and the 
southwestern regions outside the filament are 
non-overlapping regions. The 
concrete procedure is described as follows;
\begin{enumerate}
    \item We exclude the region with second moment higher than 1.5~$\kms$ as the overlapping region (Figure \ref{moments}).
    \item We define the regions of the red and blue clouds as the first moment $>-17$~$\kms$ and $<-17$~$\kms$, respectively.
    \item We average the first and second moments within the defined areas.
\end{enumerate}

From Figure \ref{moments}b and Figure \ref{spec}, the typical spectra at the region with second moment >1.5~$\kms$ have either two peaks, or are merged into a single peak. We see a dip at a typical velocity of $\sim$~$-17$~$\kms$. These suggest that the second moment >1.5~$\kms$ and the the first moment $>-17$~$\kms$ and $<-17$~$\kms$ criteria give suitable representatives for the red and blue clouds.

The velocity range for the red cloud is defined as from $V_\mathrm{representative, red}$-$dV_\mathrm{red}$ to $V_\mathrm{representative, red}$+$dV_\mathrm{red}$ 
and the velocity range for the blue cloud as from 
$V_\mathrm{representative, blue}$-$dV_\mathrm{blue}$ to $V_\mathrm{representative, blue}$+$dV_\mathrm{blue}$ where $V_\mathrm{representative, red}$, $V_\mathrm{representative, blue}$, $dV_\mathrm{red}$, $dV_\mathrm{blue}$ are
$-16.00$~$\kms$, $-18.55$~$\kms$, $0.48$~$\kms$, $0.40$~$\kms$, respectively.
\begin{figure*}
	\includegraphics[width=\textwidth]{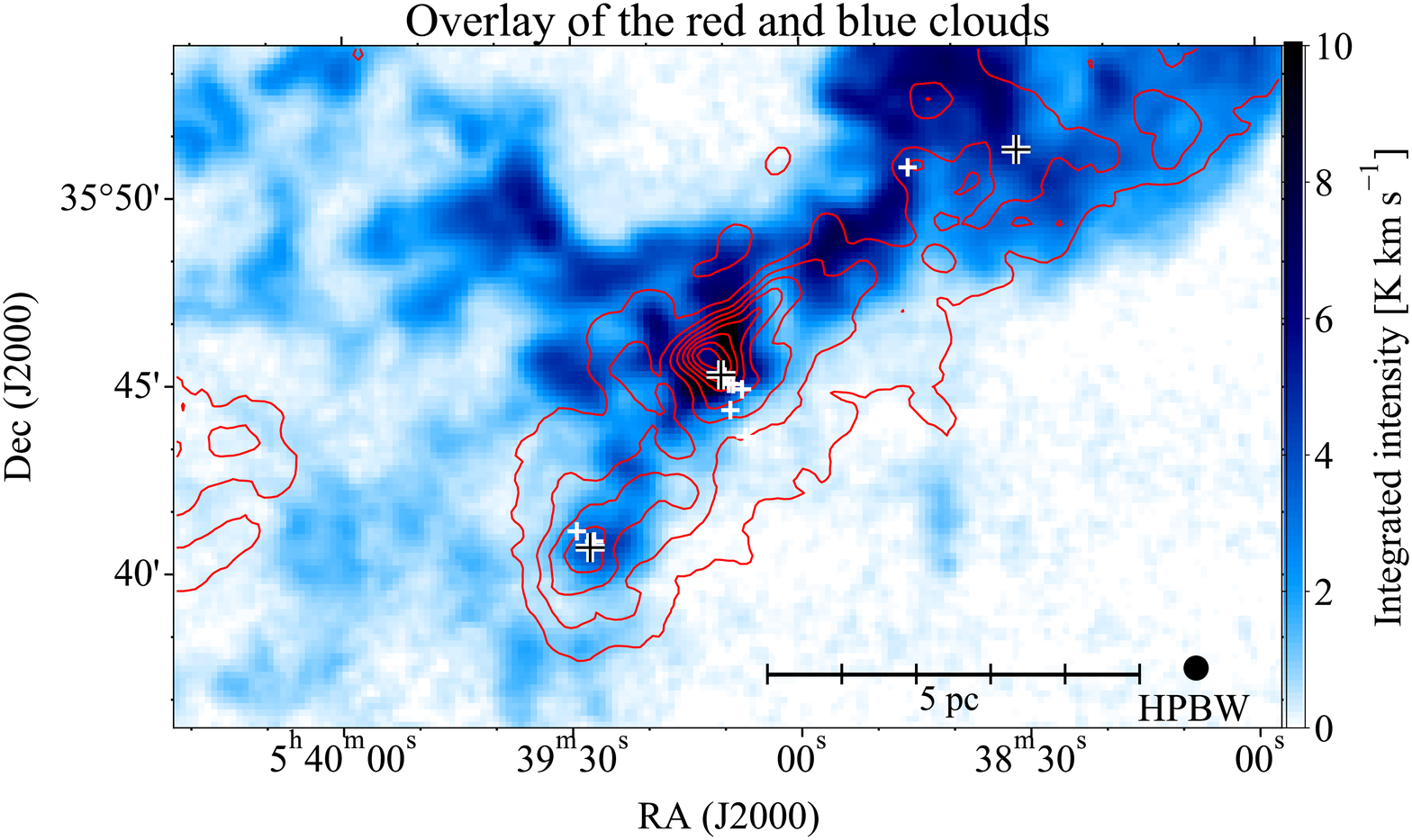}
  \caption{$\twelvecoh$ integrated intensity distribution of the red and blue clouds. The integration velocity
  range is from $-15.5$~$\kms$ to $-16.5$~$\kms$ for the image; $-19.0$~$\kms$ to $-18.2$~$\kms$ for the contours. Black crosses and white crosses are the same as Fig. \ref{Integrated_intensity}
  }
  \label{complementary}
\end{figure*}

By applying these criteria we derived the 
representative velocity range of the blue cloud 
to be $-19.0$~$\kms$ to $-18.2$~$\kms$, and that 
of the red cloud to be $-16.5$ to 
$-15.5$~$\kms$. Figure \ref{twoclouds} shows the distributions of 
the red and blue clouds, and the two 
clouds are both peaked toward IRAS~05358+3543. 
Figure \ref{complementary} shows an overlay of the two clouds, 
confirming that the filamentary distribution 
coincides with the area where the two clouds 
are overlapping. We also find that the 
blue cloud has a sharp intensity drop 
to the southwest and is extended to the northeast 
with another filamentary feature extending to the 
north from the IRAS~05358+3543 core. In contrast to the blue cloud, 
the red cloud is extended to the southeast 
with a relatively sharp drop toward the northeast.

\subsection{The physical properties of the clouds}
By assuming the local thermodynamical equilibrium we derived the column density of the molecular gas. 
For the area of the two clouds in section 3.2. for a range from $-22$~$\kms$ to $-12$~$\kms$. By assuming that the $\twelvecoh$ line is optically thick we estimated $T_\mathrm{rms}$ in each pixel. Then the $\thirteencoh$ optical depth was calculated 
from the radiation transfer equation and the $^{13}$CO column density. The ratio of H$_2$ to $^{13}$CO is assumed to 
be $5.0 \times 10^5$ \citep{Dickman1978}. The results are given in Tab. \ref{tab:physpara}. We note that the masses of the red and blue clouds include the mass of the filament. For more details, see Appendix.

\subsection{The molecular outflows associated with IRAS~05358+3543 and G173.58+2.45}

\begin{figure}
	\includegraphics[width=\columnwidth]{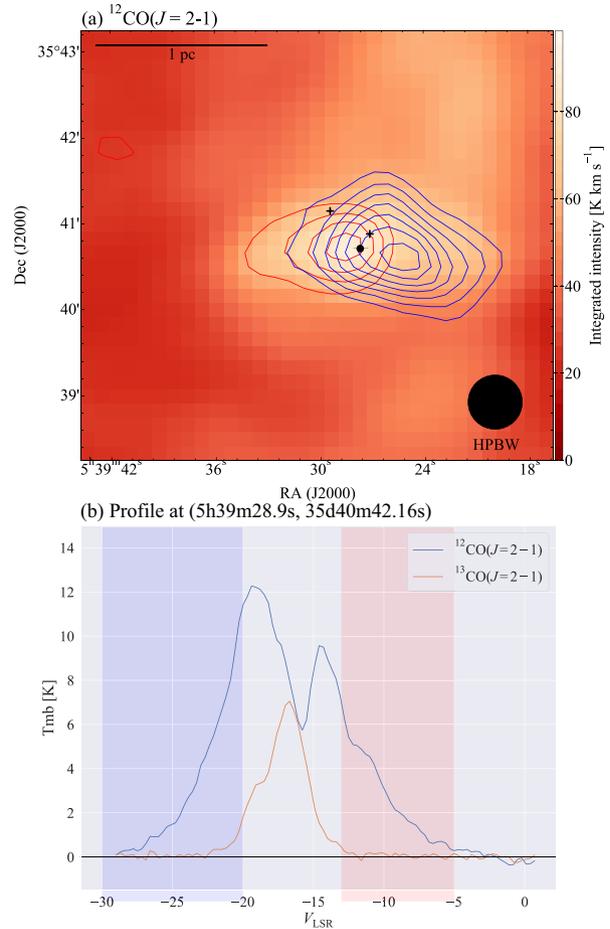}
  \caption{(a) Spatial distributions of the outflow associated with G173.58+2.45 in $\twelvecoh$. Image shows the integrated intensity distribution of the $\twelvecoh$ emission for a velocity range 
  of $-22$~$\kms$ to $-12$~$\kms$. Red and blue contours are plotted every 3 $\kkms$ from 6 $\kkms$. 
  Contour levels are selected without making a significant contamination by ambient gas. Integration velocity range is from $-30$ $\kms$ to $-21$ $\kms$ for the blue wing; from $-13$ $\kms$ to $-5$ $\kms$ for the red wing, which are shown by blue and red transparent bands in b, respectively. 
  black crosses indicate the positions of YSOs \citep{Marton2016}. (b)
  Line profile at the position shown by the black dot in (a).}
    \label{outflow_g}
\end{figure}

Two molecular outflows were detected in the $\twelvecol$ emission toward IRAS 05358+3543 and G173.58+2.45 by \cite{Snell1990} and \cite{Shepherd1996}, respectively. Figures \ref{outflow} and \ref{outflow_g} show their spatial distributions of the $\twelvecoh$ and $\thirteencoh$ outflow wings and typical profiles in the present data. The velocity ranges and the other physical parameters including size, outflow velocity, mass, dynamical time scale, mass loss rate, and outflow momentum are calculated for a detection limit of 5~$\times T_\mathrm{rms}$ under an assumption of intensity ratio of the $\twelvecoh$/$\twelvecol$ of 0.8 by referring to the $\twelvecol$ emission \citep{Shepherd1996} where the same as assumed in IRAS~05358+3543. The $X_\mathrm{CO}$ factor is assumed to be 1.0~$\times$~10$^{20}$ [$\kkms$~cm$^{-2}$] \citep{Okamoto} for the two objects. The results listed in Tab. \ref{tab:outflow} are basically consistent with the previous observations, while the size of the IRAS~05358+3543 outflow was found to be a factor of about two smaller than that in \cite{Snell1990}. This could be due to the higher resolution of the present data 38\arcsec~than that by \cite{Snell1990}, who adopted a grid spacing of 1--1.5\arcmin~with a 45\arcsec~beam.

\begin{table}
	\centering
	\caption{Physical parameters of the molecular clouds}
	\label{tab:physpara}
	\begin{tabular}{lccr} 
		\hline
		Cloud name& Column density & Mass\\
 & (cm$^{-2}$) & ($\msun$)\\
(1) & (2) & (3)\\
    \hline
    \hline
		Filament & $2.2 \times 10^{22}$ & $1030$\\
		Red cloud  & $1.3 \times 10^{22}$ & $580$\\
		Blue cloud  & $1.9 \times 10^{22}$ & $1630$\\
		\hline
  \end{tabular}
  \vspace{3pt}

         {\raggedright Notes. -- Col.(1) Cloud name. Col.(2) The peak column density of each cloud/filament. Col.(3) Masses derived by assuming the local thermodynamical equilibrium in $\alpha_\mathrm{J2000}$ and $\delta_\mathrm{J2000}$ ranges of $05^\mathrm{h}38^\mathrm{m}31^\mathrm{s}$--$05^\mathrm{h}39^\mathrm{m}12^\mathrm{s}$ and $35$\degr37\arcmin30\arcsec --$35$\degr51\arcmin46\arcsec, respectively. For the filament, we used the region enclosed by a third-lowest contour corresponding to 14~$\kkms$ in Fig. \ref{Integrated_intensity}. For the red/blue cloud, The mass is derived by the region where the first moment is larger/smaller than -17~$\kms$ and the second moment is smaller than 1.5~$\kms$ in a velocity range from $-22$~$\kms$ to $-12$~$\kms$ in Fig. \ref{moments}. For more details of the method, see \cite{Enokiya_2021}. Note that we only use voxels with intensity higher than 5 $\times T_\mathrm{rms} = 0.6$~K in Col.(2)--(3).
         \par}
\end{table}

\begin{table*}
	\caption{Physical parameters of the molecular outflows}
	\label{tab:outflow}
	\begin{tabular*}{\textwidth}{lccccccccr} 
		\hline
		Lobe name & Size & $V_\mathrm{sys}$ & $V_\mathrm{outflow}$ & Mass & $t_{\mathrm{dyn}}$ & Mass loss rate & Momentum\\
		& (pc) & ($\kms$) & ($\kms$) & ($\msun$) & (Myr) & ($\msun$~yr$^{-1}$) & ($\kms$~$\msun$)\\
		(1) & (2) & (3) & (4) & (5) & (6) & (7) & (8)\\
    \hline
    \hline
		IRAS~05358+3543 blue lobe & $0.73$ & $-16.1$ & $12.3$ & $44$& $0.14$& $3.1 \times 10^{-4}$& $541$\\
    IRAS~05358+3543 red lobe & $0.56$ & $-16.1$ & $10.5$ & $28$& $0.12$& $2.3 \times 10^{-4}$& $294$\\
    G173.58+2.45 blue lobe & $0.71$ & $-16.7$ & $12.9$ & $53$& $0.13$& $4.0 \times 10^{-4}$& $683$\\
    G173.58+2.45 red lobe & $0.98$ & $-16.7$ & $10.8$ & $48$& $0.23$& $2.0 \times 10^{-4}$& $518$\\
		\hline
  \end{tabular*}
  \vspace{3pt}
         {\raggedright Notes. --- Col. (1) Lobe name. Col. (2) Effective diameters defined by ($A$/$\pi$)$^{1/2}\times$0.5, 
         where $A$ is the region enclosed by the lowest contours in Figs. \ref{outflow}a and \ref{outflow_g}a. Col. (3) Systemic
         velocities defined as peak velocities of $\thirteencoh$. Col. (4) Three-sigma level maximum radial velocity of the outflow lobe with respect to the systemic velocity. Col. (5) Masses of the lobes. Col. (6) Dynamical time scale derived by 
         dividing the two most distant points on the lowest contour by $V_\mathrm{outflow}$ Col. (7) Mass loss rate derived by Mass/$t_\mathrm{dyn}$.
         Col. (8) Momentum estimated by Mass$\times V_\mathrm{outflow}$.
         \par}
\end{table*}

\begin{figure*}
	\includegraphics[width=\textwidth]{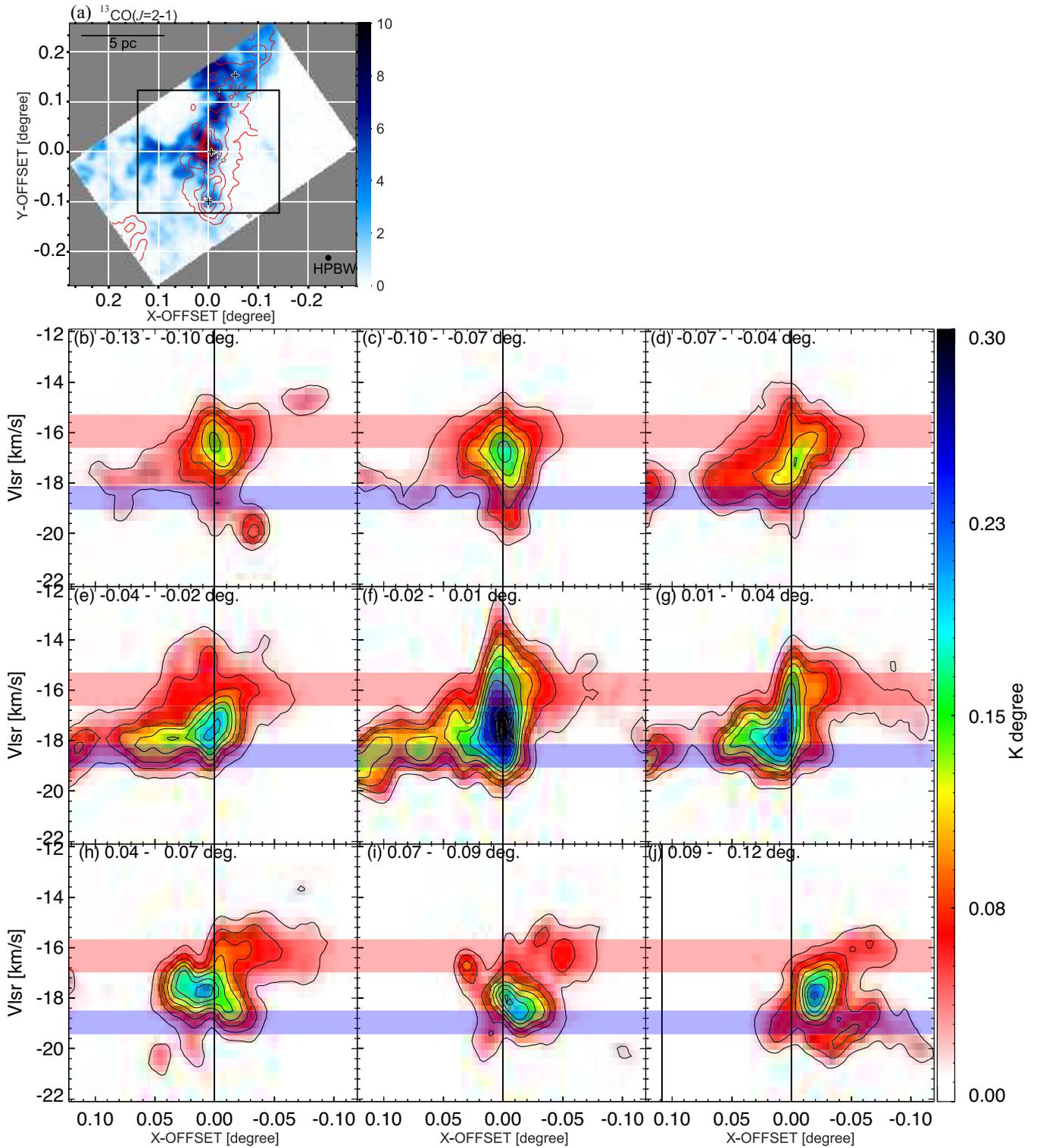}
  \caption{(a) Same as Fig. \ref{complementary} but in the X-OFFSET and Y-OFFSET coordinate. The X-OFFSET--Y-OFFSET coordinate 
  is defined by rotating the equatorial coordinate (J2000) counter clockwise by 45 degrees. The black box indicates the region
  we present X-OFFSET--velocity channel distributions in (b) to (j). Black crosses and white crosses are the same as Fig. \ref{Integrated_intensity}. (b)--(j) X-OFFSET--velocity channel distributions integrated in 
  Y-OFFSET intervals of 0.25 deg. from a Y-OFFSET $-0.13$ deg. to 0.12 deg. Contours are plotted every 0.03~K~deg. from 0.02~K~deg.}
    \label{pvchmap}
\end{figure*}

\section{Discussion}\label{sec:discussion}
\subsection{Triggered star formation in the Sh2-233 region}

\begin{figure*}
	\includegraphics[width=\textwidth]{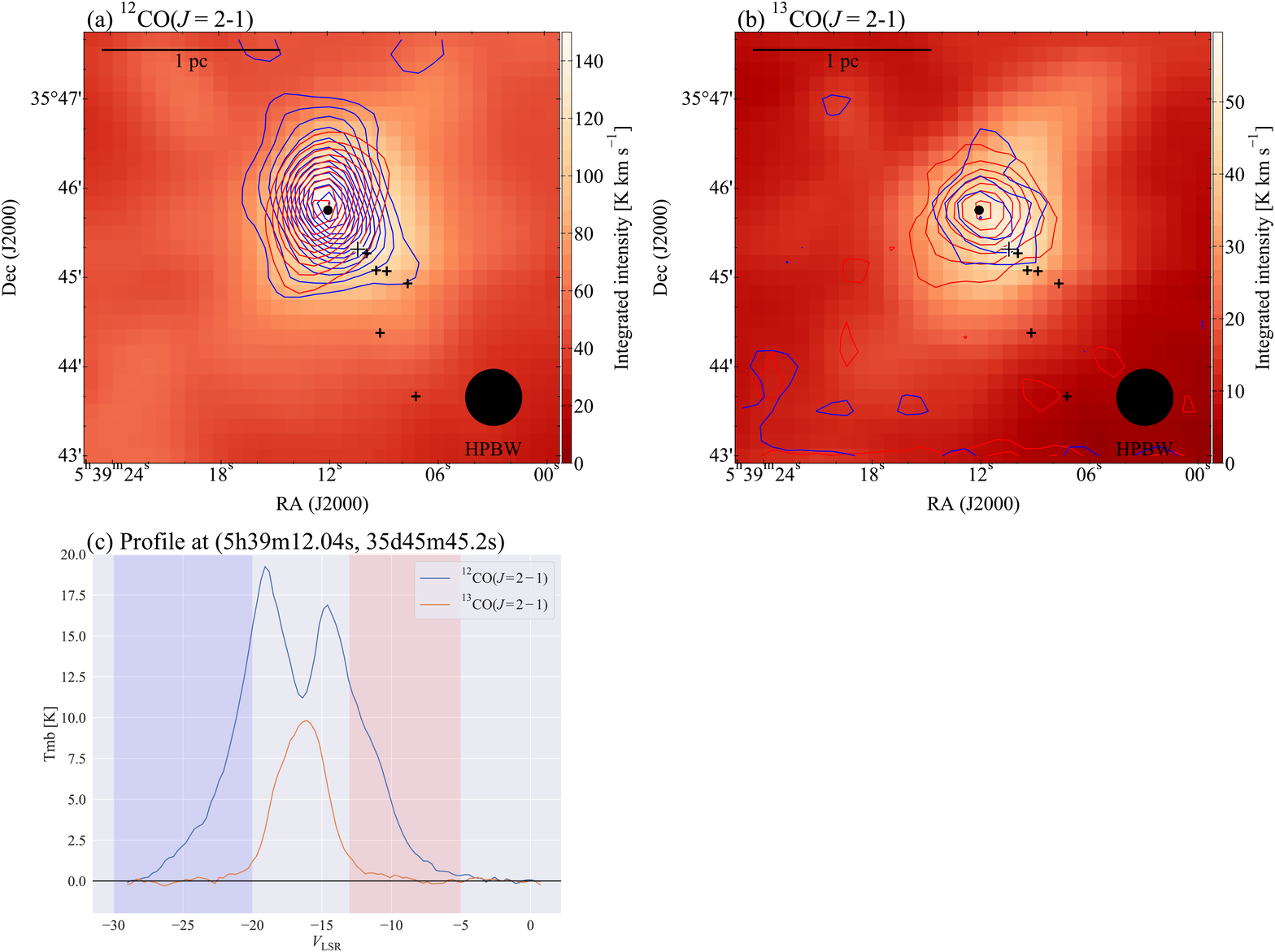}
  \caption{(a) Spatial distributions of the outflow associated with IRAS~05358+3543 in $\twelvecoh$. 
  Image shows the integrated intensity distribution of the $\twelvecoh$ emission for a velocity range 
  of $-22$~$\kms$ to $-12$~$\kms$. Red and blue contours are plotted every 2 $\kkms$ from 9 $\kkms$. 
  Contour levels are selected to avoid a significant contamination by ambient gas. Integration velocity range is from -30 $\kms$ to 
  -21 $\kms$ for the blue wing; from $-12$ $\kms$ to $-6$ $\kms$ for the red wing, which are shown by blue and red transparent 
  bands in (c), respectively. (b) Same as (a), but in $\thirteencoh$. Image shows the integrated intensity distribution of 
  the $\thirteencoh$ emission for a velocity range of $-22$~$\kms$ to $-12$~$\kms$. Red and blue contours are plotted every 
  0.5 $\kkms$ from 0.5 $\kkms$. Black dot shows the position we present typical velocity profile 
  in (c). Larger black cross and smaller black crosses indicate the positions of IRAS~05358+3543, 
  YSOs \citep{Marton2016}, respectively. (c) Line profile at the position shown by the black dot in (a).}
    \label{outflow}
\end{figure*}

\begin{figure*}
	\includegraphics[width=\textwidth]{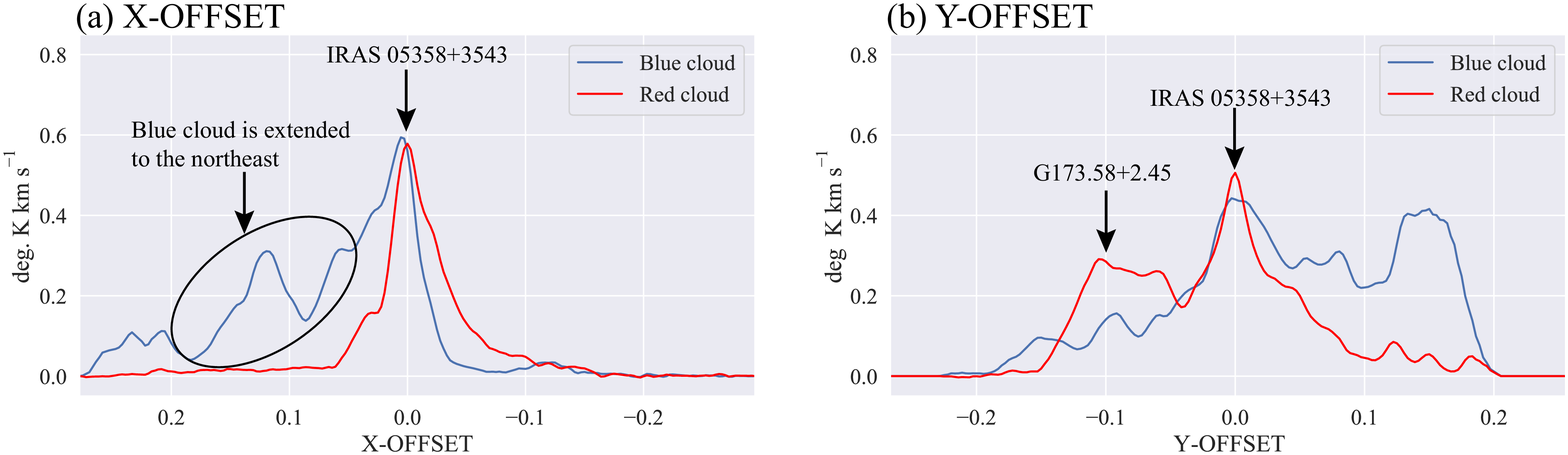}
  \caption{(a) Intensity profile in X-OFFSET integrated from $-0.05$ deg. to 0.05 deg. in Y-OFFSET on Fig. \ref{pvchmap}a.
    The profiles of the red and blue 
    clouds are plotted in red and black, respectively. (b) Same as (a) but in Y-OFFSET integrated from $-0.05$ deg. to 0.05 deg. along X-OFFSET on Fig. \ref{pvchmap}a.
    }
    \label{xyprofile}
\end{figure*}

The present Sh2-233 region is located relatively far from the other $\htwo$ regions including Sh2-235, and the present results show that the molecular gas in the region consists of two clouds at different velocities. It is possible that the two clouds, having 
complementary distribution (Fig. \ref{complementary}), are colliding with each other. If we assume that the collision took place in the northeast to the southwest direction, it is likely 
that the collision compressed the gas to form the filamentary clouds between them. 
It is also probable that the formation of the filament accompanied the formation of 
the dense cores in the filament as suggested by the magnetohydrodynamical numerical 
simulations of colliding molecular flows \citep{Inoue_Fukui}. 

Another piece of evidence for colliding clouds is found in the position velocity diagram in Figure \ref{pvchmap}, indicating a velocity jump of two clouds which are connected with each other. This distribution is understood as caused by collisional merging, where the intermediate velocity gas is caused in the interface layer where gas is mixed in velocity between the two clouds. Figure \ref{pvchmap}f shows that there are weak $\thirteencoh$ emissions outside of a velocity range of $-21$~$\kms$ to $-12$~$\kms$ localised toward X-OFFSET=0. These weak emissions are possibly ascribed to part of the outflow wings (Figure \ref{outflow}), whereas it is not likely that the weak $\thirteencoh$ outflow wings significantly alter the velocity jump.

We find that the blue cloud shows asymmetry relative to the filamentary 
cloud in the sense that the gas is extended to the northeast, while there is 
little gas distributed in the southwest. It is possible that the gas is more 
extended to the southwest of the blue cloud. The asymmetry is explained as due 
to sweeping up of the initially extended gas distribution by the red cloud, 
the compressed gas column density amounts to 2.2~$\times$~10$^{22}$~cm$^{-2}$. 
A similar, less pronounced asymmetry in the opposite sense is found in the red cloud that 
has extended gas in the southwest of the filament, which may be explained by a 
similar sweeping up on the opposite side. It is possible that the low density 
gas in the two clouds was compressed and enhanced as the filamentary cloud within 
$\sim$1~Myr, the collision time scale as shown below. So, the large-scale gas distribution 
at $\sim$5~pc seems also to be consistent with the collision picture.

The data available allow us to construct a detailed collisional picture of the formation 
of the cloud and stars. We frame a scenario of a cloud-cloud collision as follows; the 
blue cloud in the northeast collided with the red cloud about $\sim$6~$\times$~10$^5$~yr 
ago (=1.5~pc/2.5~$\kms$) by assuming that the angle between the collision velocity vector and 
the line of sight is 45 deg. The timescale varies depending on the assumed angle from 
5~$\times$~10$^5$~yr to 8~$\times$~10$^5$~yr for a range of angle from 30 deg. to 60 deg. The main range of the collision is 
from 38$^\mathrm{m}$50$^\mathrm{s}$ to 39$^\mathrm{m}$40$^\mathrm{s}$ in RA and from 
35\degr38\arcmin to 39\degr48\arcmin in Dec. The collision 
between the two clouds compressed the interface layer to form the filamentary 
cloud over $\sim$5 pc in the direction perpendicular to the collision direction. 
It seems the collision produced at least the two cores toward Sh2-233IR and 
G173.58+2.45 nearly synchronously.

\subsection{The two regions of recent star formation triggered by the cloud-cloud collision}

The most active star formation is taking place in the core toward Sh2-233IR where 
IRAS 05358+3543 is located. The core has a radius of 0.4~pc and its mass is 
estimated to be 230 $\msun$. For details, see Appendix. 
\cite{Porras2000} made {\it JHK} photometric observations of Sh2-233IR and 
identified two clusters NE and SW, which are separated with each other 
by 0.5~pc in the direction perpendicular to the filament. The NE cluster 
on the filament is significantly redder than the SE cluster, which is 
more exposed and shifted to the southwest from the filament. IRAS~05358+3543
is most likely a member of the cluster NE and the associated outflow has a 
dynamical age of $\sim$10$^4$~yr.  This dynamical age is shorter than the 
collisional time scale of (5-8)×105 yr, which is consistent with the numerical 
simulations of colliding clouds \citep{Fukui_core_mass}, 
and does not contradict the stellar age upper limit, 2 Myr. 

A question arising is if the formation of the cluster is due to the gravitational 
instability in the filamentary cloud and the gas flow along the filament collects 
the mass into the cluster. We suggest that the cluster formation is more strongly 
influenced by the initial density distribution prior to the filament formation than 
the flow along the filament, and dominated by the dynamical compression in the collision. 
This is because the column density of the blue cloud is enhanced in the northeast 
of the Sh2-233IR and by the collision the enhanced density will increase the core mass more rapidly than in the rest of the filament. Figure \ref{xyprofile} shows two strips along and perpendicular to 
the filament including the two cores and clearly shows that the column density is 
enhanced at X-OFFSET=0.0. We suggest that the enhancement of the blue cloud in column density as indicated by the cloud extended to the north favoured the formation of the Sh2-233IR core, and thereby the formation of IRAS 05358+3543. At an even smaller sub-pc scale Sh2-233IR has sub-mm/mm clumps resolved into several dust condensations mm1a-b, mm2a-d, mm3, several of which have protostellar nature \cite{Larionov,Beuther2002,Beuther2007,Leurini2007} and one of them, mm1a is likely associated with the 1.3~mm continuum emission, which is extended in the same direction with the filament at a $\sim$1\arcmin~scale. This elongation may be indicate the effect of the collision.

The G173.58+2.45 core is associated with IRAS~05361+3539 and an ultra-compact $\htwo$ region. 
CO outflow is driven by a late B or mid A star \citep{Shepherd1996}
in a small cluster of Class I and II sources \citep{Chakraborty2000,Shepherd2002, Varricatt2005}, 
H$_2$O masers \citep{Wouterloot1989, Palagi1993}, and H$_2$ emission shocks \citep{Chakraborty2000,Varricatt2005}.

Another core is toward the $\htwo$ region Sh2-233 having a B1 star
(Hunter and Massey 1990) with an age of 10$^5$ yr and is associated with IRAS 
source (IRAS05351+3549) \citep{Casoli1986,Wouterloot1989} and 
no masers \citep{Wouterloot1989,Wouterloot1993,Bronfman1996}.
\cite{Jiang2000} reported outflow toward the source, but no CO wing profile was shown. 
The core has loose connection to the present collision and requires to be confirmed.

The column densities toward the two cores are (1-2)~$\times 10^{22}$~cm$^{-2}$ at several times 0.1~pc, and they are forming at least a single high-mass star in each core. This is consistent with the statistics of other $\sim$50 cloud-cloud collision objects compiled by \cite{Enokiya_2021_compile} and \cite{Fukui_rev}, where a column density of over 10$^{23}$~cm$^{-2}$ and a collision velocity of $\sim$10~$\kms$ are the typical conditions to form more than 10 OB stars, and a column density of 10$^{22}$~cm$^{-2}$ and a collision velocity of a few $\kms$ are the conditions to form a single OB star.

\subsection{A cloud-cloud collision which provided the initial condition for high-mass star formation}

The column density above in IRAS~05358+3543 is close to 10$^{22}$ cm$^{-2}$ at 1~pc scale. It is shown that the column density becomes as high as 10$^{23}$ at 0.1 pc scale \citep{Beuther2002}. Such physical conditions, the high column density of IRAS~05358+3543, W3 Main, the ONC etc., 
as high as 1~g~cm$^{-2}$ at a sub-pc scale, have been referred to as the typical value of 
the initial mass density up to 100~$\msun$ within a 0.1~pc radius for high-mass star 
formation in the previous theoretical 
works (e.g., \citealp{Krumholz_mass_acc}). These studies of the monolithic collapse model 
assumed such high gas density in a small scale as the initial condition for 
high-mass star formation, whereas it was not explored how such extremely high 
density can be achieved. The present results indicate that a cloud-cloud collision is able to realise such a condition via strong gas compression, where the collision
may play an essential role of compression. This is consistent with the theoretical results \citep{Inoue_Fukui,Fukui_core_mass}, and suggests that a cloud-cloud collision 
is an essential process for a complete scenario of high-mass star formation which encompasses 
the formation of the massive dense core prior to the collapse. Without 
a rapid trigger it seems difficult to collect mass to such high column density by avoiding consumption of 
the gas for low-mass star formation \citep{Inoue_2018,Fukui_rev}.

\section{Conclusions}\label{sec:conclusion}
It is suggested that IRAS 05358+3543 in Sh2-233IR is a promising candidate for 
a high-mass protostar. The massive and compact dust condensation of a 0.01 pc 
radius toward IRAS 05358+3543 is considered to be a convincing signature of protostellar 
nature, whereas the formation mechanism of the condensation 
remains unexplored so far, leaving incomplete an evolutionary picture of high-mass star 
formation. In order to obtain a comprehensive evolutionary picture on IRAS 05358+3543 we 
have analysed the CO $J$=2--1 data covering more than 5~pc $\times$ 5~pc at $\sim$0.5~pc resolution 
taken with SMT in the Sh2-233 region and obtained the following results;

\begin{itemize}
  \item The new kinematical analysis of the molecular gas based on the CO($J$ = 2--1) data 
  revealed significant details of the molecular gas around IRAS 05358+3543. 
  The molecular gas shows a marked filamentary distribution of 5 pc length with 
  a 1.5~pc width and has a total mass of 1000~$\msun$. We find two $^{13}$CO peaks in the 
  filamentary cloud, where the youngest stellar objects with an age of <2 Myr are 
  associated. The most outstanding dense cloud core is located toward IRAS~05358+3543 
  and the second densest toward G173.58+2.45.

  \item The velocity field around the filament, which is elongated in the northwest to southeast
  direction, shows a significant systematic difference across the filament. The northeastern part 
  has velocity of -18.6 $\kms$ with a range of -19.0 to -18.2 $\kms$ and the southwest has 
  velocity of -16.0 $\kms$ with a range of -16.5 to -15.5 $\kms$. Their velocity difference 
  2.6 $\kms$ is significant as compared with the linewidth 1 $\kms$ in these clouds, and 
  their internal velocity field has an insignificant velocity gradient within $\sim$5\arcmin in filament. We find that 
  the two velocity components show complementary distribution with each other and the 
  filamentary cloud is located toward the boundary between the two components. 
  The filamentary cloud shows $^{13}$CO intensity significantly enhanced by a factor 
  of a few, where the two clouds are overlapped. Based on these properties, we infer 
  that a collision between the two clouds took place in the northeast-southwest 
  direction, and compressed the gas to form the filament. 
  
  \item We frame a scenario that the cloud-cloud collision triggered the formation 
  of the two dense cores in the filament and the young stellar objects therein. We 
  find a discontinuous velocity jump toward the filament, a possible trace of the 
  collisional merging. There is additional velocity broadening due to the 
  protostellar outflow in IRAS~05358+3543, which is not affecting the velocity gap significantly.
  The timescale of the collision is estimated 
  to be (5-8)~$\times$~10$^5$yr by a ratio of the filamentary  width 1.5~pc and the velocity 
  difference for a range of the assumed angle 30--60 deg. between the collision 
  velocity and the sightline.
  
  \item The present results indicate that IRAS~05358+3543 is associated with a 
  filamentary cloud, which is elongated from the southeast to the northwest. 
  We identify a dust feature possibly corresponding to this filamentary feature 
  at 1~pc scale in the continuum emission at 1.3~mm. 
  At an even smaller scale, we find a dust emission feature of 100~$\msun$ at 0.1~pc scale 
  These corresponding features suggest that the protostellar source has been formed as 
  part of the filament by the collisional compression in the present scenario. This is 
  consistent with the numerical simulations of colliding molecular flows. Further, at sub-pc scale, the sub-mm and mm continuum 
  images show dust condensations of 0.01~pc scale in IRAS~05358+3543. 

  \item On a pc scale the blue-shifted component shows a secondary filamentary 
  distribution of 5-pc in length nearly perpendicular to the filament. 
  This feature appears to cross with the filament toward the position 
  of IRAS~05358+3543. If we assume that the elongation provides more enhanced density 
  prior to the collision than elsewhere in the filament, it provides an explanation on 
  the location of the IRAS~05358+3543 core. We also note that the IRAS~05358+3543 core has two star 
  clusters, NE and SW, which are separated by 0.5~pc. 
  G173.58+2.45 is also likely formed by the same trigger in the collision, where the 
  lower density led to the formation of a less massive system with a single cluster. 
  The two objects are found to be associated with molecular outflows with dynamical timescale 
  of $\sim$10$^5$ yr, which agree with the simultaneous onset of triggering by the collision 
  separated by 3~pc. The column density of the two cores is (1--2)~$\times$~10$^{22}$ cm$^{-2}$, and values 
  which meet the criterion for the formation of a single O star triggered by a cloud-cloud collision, according to 
  the statistics of cloud-cloud collision candidates.  

  \item The physical conditions, the high column density of IRAS~05358+3543, 
  W3~Main, M42 etc., as high as 1~g~cm$^{-2}$ at a sub-pc scale, have been referred to as 
  the typical value of the initial column density for high-mass star formation in 
  the previous theoretical works. These studies of the 
  monolithic collapse model assumed such high column density in a small scale as the 
  initial condition for high-mass star formation, whereas it was not explained how such 
  extremely high density can be achieved. The present results indicate that a 
  cloud-cloud collision enables to realise such a condition, indicating that the 
  strong gas compression by a collision is essential for a comprehensive 
  scenario of high mass star formation. It is argued that without a rapid collisional trigger 
  it seems difficult to collect mass to such high column density without first consuming 
  the gas by formation of lower mass stars.

\end{itemize}


\section*{Acknowledgements}
We are grateful to Akio Taniguchi, for their valuable support during data analysis. 
We also acknowledge 
Kazuki Tokuda, Kenta Matsunaga, Mariko Sakamoto, and Takahiro Ohno 
for useful discussion of this paper. The Heinrich Hertz Submillimeter 
Telescope is operated by the Arizona Radio Observatory, a part of Steward 
Observatory at The University of Arizona. This research made use of 
Astropy,\footnote{http://www.astropy.org} a community-developed core Python 
package for Astronomy \citep{astropy:2013, astropy:2018}. 
This research made use of APLpy, an open-source plotting package for 
Python hosted at http://aplpy.github.com. This work was supported by 
JSPS KAKENHI Grant Numbers JP15H05694, JP18K13580, JP19K14758, JP19H05075, 
JP20K14520, and JP20H01945. 

\section*{DATA AVAILABILITY}
The data underlying this article will be shared on reasonable request to the corresponding author.


\bibliographystyle{mnras}
\bibliography{reference} 



\appendix

\section{LTE analyses}
We derived the column density and mass by assuming the Local thermodynamical equilibrium. We used 
voxels with intensity higher than $6\times T_\mathrm{rms}$ for $\twelvecoh$ and $\thirteencoh$. Assuming that the $\twelvecoh$ emission is optically thick, 
the excitation temperature Tex of each pixel is derived by 
\begin{equation}
    T_{\mathrm{ex}}=11.06\left\{\ln\left[1+\frac{11.06}{T_{\mathrm{peak}}+0.19}\right]\right\}^{-1}.
\end{equation}
The equivalent brightness temperature $J(T)$ is obtained by
\begin{equation}
    J(T)=\frac{h\nu}{k_\mathrm{B}}\left[\exp{\left(\frac{h\nu}{k_{\mathrm{B}}T}\right)}-1\right]^{-1},
\end{equation}
where $h$, $k_\mathrm{b}$, $\nu$ are the Plank constant, Boltzmann constant, and observing frequency. From the radiation transfer equation, 
the optical depth($\tau$) of the $\thirteencoh$ emission is given by
\begin{equation}
    \tau(v)=-\ln{\left[1-\frac{T_{\mathrm{mb}}}{J(T_{\mathrm{ex}})-J(T_{\mathrm{bg}})}\right]}.
\end{equation}
Using $\tau$, the column density of $\thirteenco$ is shown by the following equation:
\begin{equation}
N=\sum_v \tau(v)\Delta v \frac{3k_\mathrm{B}T_{\mathrm{ex}}}{4\pi^3\nu\mu^2}\exp{\left(\frac{h\nu J}{2k_{\mathrm{B}}T_{\mathrm{ex}}}\right)}\times \frac{1}{1-\exp{(-h\nu/k_{\mathrm{B}}T_{\mathrm{ex}})}}.
\end{equation}
Substituting $k_\mathrm{B}$ = $1.38 \times 10^{-16}$ (erg K$^{-1}$), $\nu = 2.20\times10^{11}$~(Hz), 
$\mu = 1.10 \times 10^{-19}$~(esu cm), $h = 6.63 \times 10^{-27}$~(erg s), $J = 1$, and $T_\mathrm{ex}$ values of each pixel, 
we get the column density of H$_2$ molecule by assuming an abundance ratio to H$_2$ of $\thirteenco$ as $5\times10^5$ \citep{Dickman1978}. 
The mass of molecular clouds is described as
\begin{equation}
M = m_{\mathrm{p}} \mu_\mathrm{m} D^2\Omega \sum_{i}N_{i}(\mathrm{H_2}),
\label{mass}
\end{equation}
where $\mu_m$, $m_\mathrm{p}$, $D$, $\Omega$, and $N_{i}(\mathrm{H_2})$ are mean molecular weight, proton mass, 
distance, solid angle subtending a size of the pixel, and the column density of molecular hydrogen for the $i$th pixel, respectively. 
We assume the Helium abundance of 20\%, corresponding to a mean molecular weight of 2.8, 
and a distance of 1.8~kpc.

\section{Deriving physical parameters of the molecular cores}
\begin{figure*}
	\includegraphics[width=\textwidth]{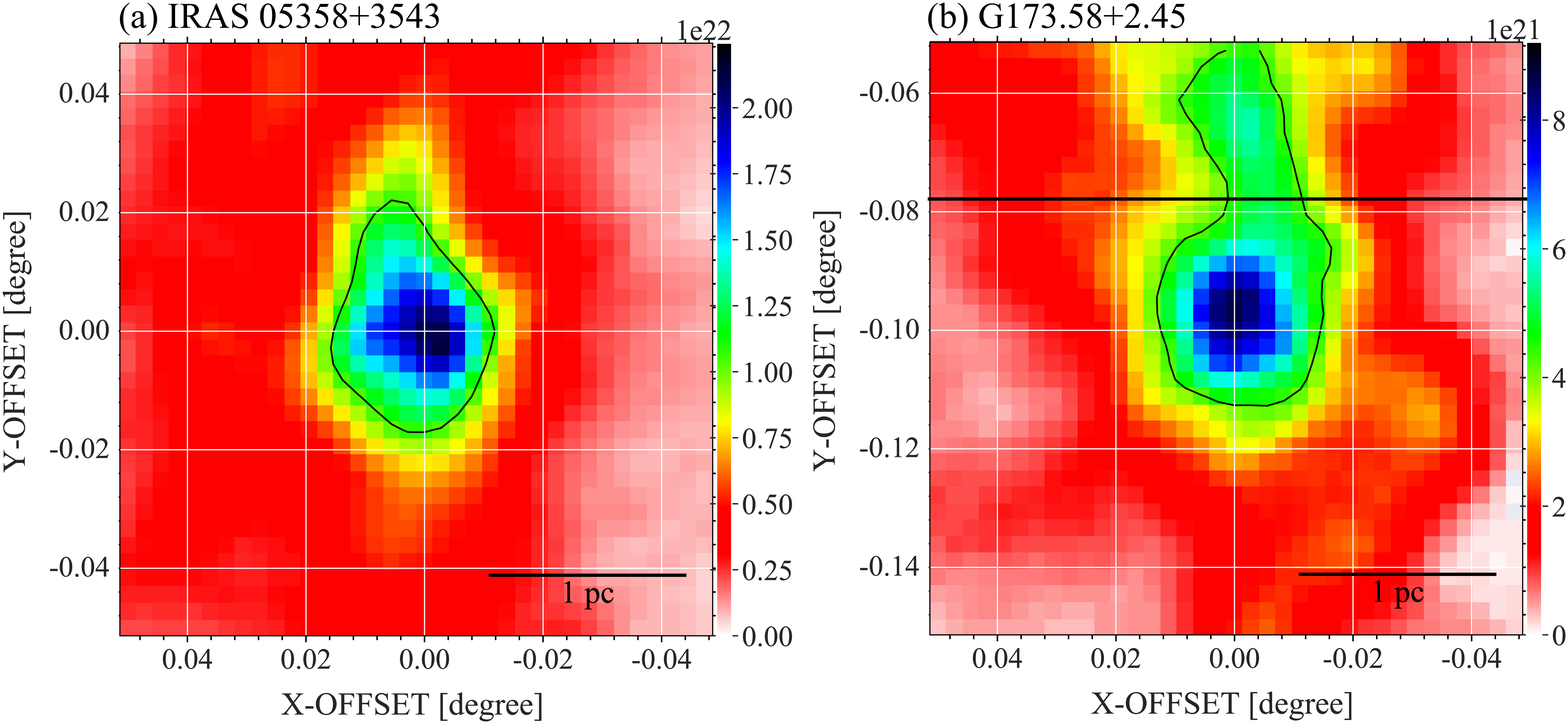}
  \caption{Close-up view of the column density map of IRAS 05358+3543. Superposed contour indicates the half of 
  the peak column density. (b)Same as (a), but for the G173.58+2.45 core. Superposed contour indicates the half of 
  the peak column density. Black horizontal line indicates a Y-OFFSET of -0.076 degree, corresponding to the intensity dip of the 
  Y-OFFSET-intensity profile at X-OFFSET=0.}
    \label{core}
\end{figure*}

Figs. \ref{core}a and b show closed-up column density maps toward IRAS 05358+3543 and G173.58+2.45 with masses of 
230~$\msun$ and 100~$\msun$, respectively. The overlaid contour indicates the half of the peak column density, and we 
derived the enclosed mass as the mass of the IRAS 05358+3543 core. The G173.58+2.45 core is elongated along 
the Y-OFFSET axis with a dip in Y-OFFSET-column density profile at Y-OFFSET = -0\fdg076. To exclude another peak 
at (X-OFFSET, Y-OFFSET)=(0\fdg0,-0\fdg06), we draw a horizontal line at Y-OFFSET = -0\fdg076, and derived the enclosed 
mass to obtain the mass of the core. 

\bsp	
\label{lastpage}
\end{document}